\documentclass[preprint,12pt]{elsarticle}
\usepackage[margin=20mm]{geometry}
\biboptions{sort&compress}
\usepackage[T1]{fontenc}
\usepackage{ragged2e}
\usepackage{amsmath,amssymb,amsfonts,mathtools,bm}
\usepackage{amsthm}
\usepackage{xfrac}
\usepackage{mathrsfs}
\usepackage{dsfont}
\usepackage{dashbox}
\usepackage{anyfontsize}
\usepackage{graphicx}
\usepackage{xcolor}
    \definecolor{notecolor}{HTML}{300d66}
    \definecolor{tmpcolor}{HTML}{AA0033}
\usepackage[colorlinks=true]{hyperref}
\usepackage{comment}
\usepackage{refcount}
\newtheorem{theorem}{Theorem}

\theoremstyle{definition}

\theoremstyle{remark}

\newcommand{\appref}[1]{\hyperref[#1]{Appendix~\ref*{#1}}}
\newcommand{\secref}[1]{\hyperref[#1]{Section~\ref*{#1}}}
\newcommand{\subsecref}[1]{\hyperref[#1]{Section~\ref*{#1}}}

\newcommand{\Hir}{M}

\newcommand{\ba}{\begin{eqnarray}}
\newcommand{\ea}{\end{eqnarray}}

\begin{document}
\hypersetup{
    linkcolor=blue,
    citecolor=blue,
    urlcolor=blue
}
\begin{frontmatter}

\title{Painlevé VI and orthogonal polynomials with discontinuous weights in stationary axisymmetric gravity}
\author[1]{Adriano de Medeiros da Silva}
\ead{ademedeirosdas@gmail.com}

\author[2]{A. R. Melikyan}
\ead{amelik@gmail.com}

\affiliation[1]{
    organization={Departamento de Matemática, Instituto de Ciências Exatas,\\Universidade de Brasília},
    city={Brasília},
    state={DF},
    postcode={70910-900},
    country={Brazil}
}

\affiliation[2]{
    organization={Centro Internacional de Física, Instituto de Física,\\Universidade de Brasília},
    city={Brasília},
    state={DF},
    postcode={70910-900},
    country={Brazil}
}

\begin{abstract}
We investigate stationary axisymmetric vacuum solutions of Einstein's equations and their connection to Painlev\'e VI. We show that Hankel blocks appearing in determinant solutions of the Ernst equation are related to orthogonal polynomials with discontinuous weights, and that successive Christoffel transformations lead to a hierarchy of associated Painlev\'e VI equations.
\end{abstract}

\end{frontmatter}


\section{Introduction}

Stationary axisymmetric vacuum solutions of Einstein's field equations can be described by the metric:
\begin{equation}\label{metric}
ds^2=f(dt-\omega\,d\phi)^2-r^2f^{-1}d\phi^2-e^\mu(dr^2+dz^2).
\end{equation}
Here \((r,\phi,z)\) are cylindrical coordinates, and the metric functions \(f\), \(\omega\), and \(\mu\) depend only on \(r\) and \(z\). The corresponding field equations reduce to the Ernst equation for the complex potential \(\xi\):
\begin{equation}\label{ernst_equation}
(\xi\xi^*-1)\nabla^2\xi=2\xi^*(\nabla\xi)^2.
\end{equation}
The Tomimatsu-Sato solutions \cite{Tomimatsu-Sato-Original}, defined for positive integer values of \(\delta\), are rational solutions of the Ernst equation of the form:
\begin{equation}
\xi_\delta(x,y;p,q)=\frac{\alpha_\delta(x,y;p,q)}{\beta_\delta(x,y;p,q)},
\end{equation}
where \(\alpha_\delta\) and \(\beta_\delta\) are complex polynomials, the variables \((x,y)\) are prolate spheroidal coordinates related to the cylindrical coordinates by \(r=k\sqrt{(x^2-1)(1-y^2)}\) and \(z=kxy\), where \(k\) is a constant, and \(p^2+q^2=1\). The generalization of the Tomimatsu-Sato solutions to arbitrary positive integer values of \(\delta\) was proposed by Yamazaki and Hori in \cite{Yamazaki:1977xd} and proved by Hori in \cite{Hori:1978ie}.

Cosgrove considered in \cite{Cosgrove:1977np} the problem for continuous values of \(\delta\) and constructed a family of exact solutions containing the known Kerr and Tomimatsu-Sato metrics for \(\delta=1,2,3,4\) (see also \cite{CMCosgrove_1978,CMCosgrove_1978_2}). His construction expresses the metric and Ernst potentials in terms of solutions of two second-order ordinary differential equations.
Introducing the coordinates
\begin{equation}
\eta=\frac{x^2-1}{1-y^2},
\qquad
\nu=\frac{y}{x},
\end{equation}
Cosgrove obtained a nonlinear equation for \(H_4(\eta)\) (see Eq.~\eqref{H_4}), which can be transformed into the Jimbo-Miwa-Okamoto \(\sigma\)-form of Painlev\'e VI \cite{CosgroveScoufis1993,Okamoto1986}.
 The second is a linear Fuchsian equation for \(K^{(\epsilon)}(\nu,\eta)\), whose coefficients depend on \(H_4(\eta)\) and its derivatives, with \(\eta\) treated as a parameter.
 The equation for \(K^{(\epsilon)}\) is obtained by reducing an auxiliary Appell equation to a second-order linear equation. Once a solution of the equation for \(H_4(\eta)\) is known, the coefficients of this linear equation can be determined. Solving it with the appropriate boundary conditions then allows the reconstruction of the Ernst potential and the corresponding metric functions. For integer \(\delta\), the resulting solutions are rational or polynomial, while noninteger values generally lead to transcendental functions. Cosgrove also introduced an additional parameter \(h\) into both differential equations and relaxed the boundary conditions, obtaining a larger family of solutions that contains the original asymptotically flat family when \(h=0\). In the Painlev\'e VI formulation, this parameter enters through the parameters of the nonlinear equation for \(H_4\).
 Dale \cite{Dale:1978xx} subsequently obtained determinant solutions of Cosgrove's \(H_4\) equation for \(h=0\).

In this paper,\footnote{This work is partly based on \cite{AdrianoTCC}.} we investigate the relation between Dale's determinant solutions of the \(H_4\) equation and the determinant representations of the Yamazaki-Hori solutions of the Ernst equation. To this end, we use the equivalent determinant representation obtained in \cite{Melikyan:2025rmj}, which involves an arbitrary parameter \(\kappa\). For a suitable choice of \(\kappa\), we show that the Hankel matrices appearing in this representation can be expressed in terms of moments of discontinuous weights.
In particular, the original Hankel block is associated with the same weight as Dale's determinant. This allows us to use the results of \cite{Chen:2010rl,Min:2020ld} on orthogonal polynomials with discontinuous weights to identify the corresponding Painlev\'e VI parameters. We then extend this connection by considering Christoffel transformations of the Yamazaki-Hori matrices and identifying the corresponding Painlev\'e VI parameters, including those associated with Cosgrove's generalized equation for \(h\neq 0\).

The paper is organized as follows. In Section \ref{section2}, we use Cosgrove's asymptotic condition to determine the constant \(c=-q^2/p^2\) in Dale's determinant solution. In Section \ref{section3}, we express Dale's determinant in terms of a discontinuous weight and establish the correspondence between the parameters of Painlev\'e VI and Cosgrove's generalized \(H_4\) equation. In Section \ref{section4}, we examine the original and Christoffel-transformed Hankel blocks of the Yamazaki-Hori determinants, establish their connection with Dale's determinant, and derive a relation between the corresponding Painlev\'e VI \(\sigma\)-functions. Finally, in Section \ref{section5}, we consider successive Christoffel transformations, derive the hierarchy of Painlev\'e VI parameters, and establish relations between the \(\sigma\)-functions at different levels of the hierarchy.

\section{Asymptotic limit of the $H_4$ equation and Dale's solution: $h=0$}
\label{section2}

The nonlinear differential equation for \(H_4(\eta)\) plays a central role in Cosgrove's construction of stationary axisymmetric vacuum solutions \cite{Cosgrove:1977np}. Cosgrove also generalized this equation by introducing an additional parameter \(h\):
\begin{equation}\label{H_4}
\eta^2(1+\eta)^2(H_4'')^2=4(\eta H_4'^2-H_4H_4'-h^2)[-\delta^2+H_4-(1+\eta)H_4'].
\end{equation}
We first consider the original case \(h=0\). Following \cite{Cosgrove:1977np}, we require the gravitational potentials to have the asymptotic behavior of a rotating source with mass \(m>0\), angular momentum \(J\), vanishing mass dipole moment, and mass quadrupole moment \(Q\). Introducing \(u=\frac12\ln f\) and the potential \(\psi\), defined by:
\begin{equation}
\frac{\partial\psi}{\partial r}=\frac{f^2}{r}\frac{\partial\omega}{\partial z},
\qquad
\frac{\partial\psi}{\partial z}=-\frac{f^2}{r}\frac{\partial\omega}{\partial r},
\end{equation}
the asymptotic conditions are:
\begin{align*}
u &= -m \rho^{-1} + \left( Q - \frac{1}{3} m^3 \right) \left( \frac{3}{2} \cos^2 \theta - \frac{1}{2} \right) \rho^{-3} + \mathrm{O}(\rho^{-4}), \\
\omega &= -(2J \sin^2 \theta) \rho^{-1} + \mathrm{O}(\rho^{-2}), \\
\psi &= -(2J \cos \theta) \rho^{-2} + \mathrm{O}(\rho^{-3}),
\end{align*}
as \(\rho\rightarrow\infty\), where \(r=\rho\sin\theta\) and \(z=\rho\cos\theta\). These asymptotic conditions describe the gravitational field far from the source in terms of its lowest multipole moments. The corresponding boundary condition for \(H_4(\eta)\) is:
\begin{equation}\label{Boundary_H_4}
H_4(\eta)=\frac{\delta^2}{p^2}+O(\eta^{-1}),\qquad \eta\rightarrow\infty.
\end{equation}
As shown in \cite{Cosgrove:1977np}, this condition leads to the required asymptotic behavior of the gravitational potentials.

Dale constructed in \cite{Dale:1978xx} explicit rational solutions of a differential equation equivalent to \eqref{H_4} for \(h=0\), using ratios of determinants. He proved the following theorem:

\begin{theorem}
\label{theorem1}
Let \(n\) be a positive integer and \(c\) an arbitrary constant. Define
\begin{equation}\label{Dale_matrix}
A_n=\det(a_{ij})_{i,j=1}^n,\qquad
a_{ij}=\frac{z^{i+j-1}+(-1)^{i+j}c}{i+j-1},
\end{equation}
and let \(A_{11}\) denote the cofactor of \(a_{11}\). Then, for \(a=-4n^2\),
\begin{equation}
\zeta^{(D)}(z)=4z(c-1)\frac{A_{11}}{A_n}
\end{equation}
is a solution of the differential equation
\begin{equation}\label{Dale_eq1}
z^2(1+z)^2\left(\frac{d^2\zeta^{(D)}}{dz^2}\right)^2
=
\frac{d\zeta^{(D)}}{dz}
\left(z\frac{d\zeta^{(D)}}{dz}-\zeta^{(D)}\right)
\left((1+z)\frac{d\zeta^{(D)}}{dz}-\zeta^{(D)}+a\right).
\end{equation}
\end{theorem}
\noindent For \(h=0\), Dale's equation \eqref{Dale_eq1} is equivalent to \eqref{H_4} under the identifications:
\begin{equation}\label{Dale-H_4}
z=\eta,\qquad \zeta^{(D)}(z)=-4H_4(\eta),\qquad a=-4\delta^2.
\end{equation}
We first consider the limit \(z\rightarrow\infty\). From \eqref{Dale_matrix}, we have:
\begin{equation}
a_{ij}(z,c)\sim\frac{z^{i+j-1}}{i+j-1}=z^{i+j-1}(\mathcal H_n)_{ij},
\end{equation}
where \(\mathcal H_n\) is the Hilbert matrix with entries \((\mathcal H_n)_{ij}=1/(i+j-1)\).
Factoring \(z^i\) from each row and \(z^{j-1}\) from each column, we obtain:
\begin{equation}
A_n(z,c)\sim z^{n^2}\det\mathcal H_n.
\end{equation}
For the cofactor \(A_{11}\), the row and column indices run from \(2\) to \(n\), giving a total power of \(n^2-1\). Thus,
\begin{equation}
A_{11}(z,c)\sim z^{n^2-1}\operatorname{cof}_{11}(\mathcal H_n)=n^2z^{n^2-1}\det\mathcal H_n,
\end{equation}
where we used \((\mathcal H_n^{-1})_{11}=n^2\) \cite{Choi:1983ng}. Using the asymptotic expressions for \(A_n\) and \(A_{11}\), we find for Dale's solution:
\begin{equation}
\lim_{z\to\infty}\zeta^{(D)}(z)=4n^2(c-1).
\end{equation}
For \(\delta=n\), the boundary condition \eqref{Boundary_H_4}, together with the relation \eqref{Dale-H_4}, gives:
\begin{equation}\label{asymptotic limit}
\lim_{z\to\infty}\zeta^{(D)}(z)=-4\frac{n^2}{p^2}.
\end{equation}
Comparing the two limits, we obtain:
\begin{equation}
c=1-\frac{1}{p^2}=-\frac{q^2}{p^2}.
\end{equation}

The relation \(c=-q^2/p^2\) explains the assumption \(c<0\) made by Dale in \cite{PDale_1979}, where he proved that \(A_n\) has exactly \(n\) distinct positive real zeros, corresponding to \(2n\) infinite red-shift surfaces. For real \(p\) and \(q\), with \(p\neq 0\), we have \(c\leq 0\), with strict inequality for \(q\neq 0\). The static limit \(q\rightarrow 0\) gives \(c\rightarrow 0\), while the extreme limit \(p\rightarrow 0\) gives \(c\rightarrow-\infty\).

Manko and Moreno \cite{Manko1997ExtensionOT} extended the Tomimatsu-Sato solutions to the super-extreme case by considering the analytic continuation:
\begin{equation}\label{continuation_Manko}
p\mapsto i\tilde{p},\qquad k\mapsto ik,
\end{equation}
under which the prolate spheroidal coordinates become oblate. Since \(p^2+q^2=1\), this continuation gives \(q^2=1+\tilde{p}^2\), and therefore:
\begin{equation}
c=-\frac{q^2}{p^2}=1+\frac{1}{\tilde{p}^2}>1.
\end{equation}
Thus, the sub-extreme and super-extreme cases correspond to \(c<0\) and \(c>1\), respectively.

\section{Orthogonal polynomials with discontinuous weights} \label{section3}

The orthogonal polynomials and Hankel determinants studied in \cite{Min:2020ld} are associated with the discontinuous Jacobi weight:
\begin{equation}\label{Chen_weight_again}
w(x,t)=x^\alpha(1-x)^\beta|x-t|^\gamma(A+B\theta(x-t)),
\qquad x,t\in[0,1],
\end{equation}
where \(\theta\) is the Heaviside step function, \(\alpha,\beta,\gamma>0\), and \(A,B\) are real constants satisfying \(A\geq0\) and \(A+B\geq0\). The corresponding Hankel determinant is:
\begin{equation}\label{Hankel_again}
D^{(\gamma)}_{\Hir}(t;A,B)=\det\left(\int_0^1 x^{i+j}w(x,t)\,dx\right)_{i,j=0}^{\Hir-1},
\end{equation}
where \(\Hir\) is the order of the determinant and the superscript \(\gamma\) denotes the exponent appearing in the weight. Using ladder operators for the associated orthogonal polynomials and their compatibility conditions, it was shown in \cite{Min:2020ld} that a suitable shift of the logarithmic derivative of this determinant satisfies the Jimbo-Miwa-Okamoto \(\sigma\)-form of Painlev\'e VI.

The main result of \cite{Min:2020ld} is the following theorem.
\begin{theorem}
Let \(D_{\Hir}^{(\gamma)}(t;A,B)\) be the Hankel determinant defined in \eqref{Hankel_again}, and set:
\begin{equation}\label{H_n_again}
H_{\Hir}^{(\gamma)}(t)=t(t-1)\frac{d}{dt}\ln D_{\Hir}^{(\gamma)}(t;A,B).
\end{equation}
Define:
\begin{equation}\label{sigma_again}
\sigma_{\Hir}(t)=H_{\Hir}^{(\gamma)}(t)+c_1t+c_2,
\end{equation}
where
\begin{equation}\label{c1_again}
c_1=-\Hir(\Hir+\alpha+\beta+\gamma)-\frac{(\alpha+\beta)^2}{4},
\end{equation}
and
\begin{equation}\label{c2_again}
c_2=\frac14\left[2\Hir(\Hir+\alpha+\beta+\gamma)+(\alpha+\beta)\beta-(\alpha-\beta)\gamma\right].
\end{equation}
Then \(\sigma_{\Hir}(t)\) satisfies the Jimbo-Miwa-Okamoto \(\sigma\)-form of Painlev\'e VI:
\begin{equation}\label{sigma_n_equation}
\sigma_{\Hir}'[t(t-1)\sigma_{\Hir}'']^2+\left[2\sigma_{\Hir}'(t\sigma_{\Hir}'-\sigma_{\Hir})-(\sigma_{\Hir}')^2-\nu_1\nu_2\nu_3\nu_4\right]^2
=(\sigma_{\Hir}'+\nu_1^2)(\sigma_{\Hir}'+\nu_2^2)(\sigma_{\Hir}'+\nu_3^2)(\sigma_{\Hir}'+\nu_4^2),
\end{equation}
with parameters:
\begin{equation}\label{parameters}
\nu_1=\frac{\alpha+\beta}{2},\qquad
\nu_2=\frac{\beta-\alpha}{2},\qquad
\nu_3=\frac{2\Hir+\alpha+\beta}{2},\qquad
\nu_4=\frac{2\Hir+\alpha+\beta+2\gamma}{2}.
\end{equation}
\end{theorem}

The theorem above assumes \(\alpha,\beta>0\) to ensure that the weight vanishes at the endpoints, so that no boundary terms arise when integrating by parts. However, in what follows, we consider the case \(\alpha=\beta=0\). For \(0<t<1\), the moments and their derivatives with respect to \(t\) have well-defined limits as \(\alpha,\beta\rightarrow0^+\). As a consequence, the corresponding \(\sigma\)-function continues to satisfy the Painlev\'e VI equation \eqref{sigma_n_equation} in this limit, wherever the limiting Hankel determinant is nonzero. We also consider the case \(\gamma=0\), for which the factor \(|x-t|^\gamma\) disappears and the weight reduces to the discontinuous Jacobi weight studied in \cite{Chen:2010rl}. Setting \(\gamma=0\) in the theorem above reproduces the corresponding Painlev\'e VI equation and its parameters obtained in \cite{Chen:2010rl}. This case will be used below for Dale's determinant.

\subsection{Dale's solution and Painlev\'e VI}

In Section \ref{section2}, we identified Dale's variable \(z\) with Cosgrove's coordinate \(\eta=(x^2-1)/(1-y^2)\). Since \(x>1\) and \(-1<y<1\) in the prolate spheroidal coordinate domain, we have \(z>0\). We also recall two identities for Dale's determinant \cite{Dale:1978xx}, defined in Theorem \ref{theorem1}:
\begin{align}
A_n(z,c) &= z^{n^2}c^n A_n(z^{-1},c^{-1}),\\
A_n(z,c) &= (-1)^n A_n(-1-z,1-c).
\end{align}
To express Dale's determinant in terms of moments on the interval \([0,1]\), we introduce the variable:\footnote{In what follows, derivatives with respect to \(t\) are taken at fixed \(\nu=y/x\) and fixed \(p,q\), in accordance with Cosgrove's choice of coordinates \((\eta,\nu)\).
}
\begin{equation}
t:=\frac{1}{1+z},\qquad 0<t<1.\label{t_def}
\end{equation}
The matrix elements defined in \eqref{Dale_matrix} can then be written as:
\begin{equation}
a_{ij}=(1+z)\int_0^1\bigl((1+z)\xi-1\bigr)^{i+j-2}w_D(\xi,t;c)\,d\xi,
\end{equation}
where
\begin{equation}\label{Dale's_Weight}
w_D(\xi,t;c)=c+(1-c)\theta(\xi-t).
\end{equation}
Indeed, splitting the integral at \(\xi=t\) and substituting \(u=(1+z)\xi-1\), we obtain
\begin{equation}
\begin{aligned}
a_{ij}
&=c\int_{-1}^0u^{i+j-2}\,du+\int_0^zu^{i+j-2}\,du\\
&=\frac{z^{i+j-1}+(-1)^{i+j}c}{i+j-1},
\end{aligned}
\end{equation}
which reproduces the matrix elements of \eqref{Dale_matrix}.

We define the Hankel determinant associated with Dale's weight by
\begin{equation}
D_n^{(D)}(t,c):=\det\left(\int_0^1\xi^{i+j}w_D(\xi,t;c)\,d\xi\right)_{i,j=0}^{n-1}.
\end{equation}
To relate this determinant to \(A_n(z,c)\), we expand
\begin{equation}
\bigl((1+z)\xi-1\bigr)^{i-1}=\sum_{k=1}^i\binom{i-1}{k-1}(1+z)^{k-1}(-1)^{i-k}\xi^{k-1}.
\end{equation}
This defines a lower triangular matrix \(L_n\):
\begin{equation}
(L_n)_{ik}=
\begin{cases}
\displaystyle\binom{i-1}{k-1}(1+z)^{k-1}(-1)^{i-k}, & k\leq i,\\
0, & k>i.
\end{cases}
\end{equation}
 Denoting by \(M_n(t,c)\) the Hankel matrix whose determinant is \(D_n^{(D)}(t,c)\), we obtain:
\begin{equation}
(a_{ij})_{i,j=1}^n=(1+z)L_nM_n(t,c)L_n^T.
\end{equation}
Since \(L_n\) is lower triangular, we have:
\begin{equation}
\det L_n=\prod_{k=1}^n(1+z)^{k-1}=(1+z)^{n(n-1)/2}.
\end{equation}
Thus:
\begin{equation}
A_n(z,c)=(1+z)^{n^2}D_n^{(D)}(t,c).
\end{equation}

Dale \cite{Dale:1978xx} also obtained the following expression for his solution:
\begin{equation}\label{DALE_49}
\zeta_n^{(D)}(z)=4z\left[n^2-(1+z)\frac{d}{dz}\ln|A_n(z,c)|\right].
\end{equation}
Using the relation between \(A_n\) and \(D_n^{(D)}\), together with \(t=1/(1+z)\), we obtain:
\begin{equation}\label{Dale_Hankel}
\zeta_n^{(D)}(z)=4(1-t)\frac{d}{dt}\ln|D_n^{(D)}(t,c)|.
\end{equation}
Comparing Dale's weight \eqref{Dale's_Weight} with \eqref{Chen_weight_again}, we identify
\begin{equation}
\Hir=n,\qquad \alpha=\beta=\gamma=0,\qquad A=c,\qquad B=1-c.
\end{equation}
For real \(p,q\) with \(pq\neq0\), we have \(c=-q^2/p^2<0\). Dale's weight is then negative for \(0<\xi<t\) and positive for \(t<\xi<1\), so the result of \cite{Chen:2010rl} cannot be applied directly. However, under the analytic continuation \eqref{continuation_Manko}, we obtain \(c>1\), and the weight becomes positive throughout the interval. The Painlev\'e VI equation established in \cite{Chen:2010rl} therefore holds for the continued determinant. We can extend this result to the original values \(c<0\) by observing that the moments of Dale's weight are
\begin{equation}
\mu_k(t,c)=\frac{1+(c-1)t^{k+1}}{k+1}.
\end{equation}
Thus, the Hankel determinant and its derivatives with respect to \(t\) are polynomials in \(c\). The Painlev\'e VI equation, expressed through the logarithmic derivative of the determinant, is therefore an identity between rational functions of \(c\). Since this identity holds for every \(c>1\), it also holds for \(c<0\), wherever \(D_n^{(D)}(t,c)\neq0\). The corresponding Painlev\'e VI parameters are
\begin{equation}
(\nu_1,\nu_2,\nu_3,\nu_4)=(0,0,n,n).
\end{equation}
The constants in \eqref{c1_again} and \eqref{c2_again} reduce to \(c_1=-n^2\) and \(c_2=n^2/2\). Hence, the corresponding \(\sigma\)-function is:
\begin{equation}\label{sigma_Dale}
\sigma_n^{(D)}(t)=t(t-1)\frac{d}{dt}\ln|D_n^{(D)}(t,c)|-n^2t+\frac{n^2}{2}.
\end{equation}
Combining this expression with \eqref{Dale_Hankel}, we obtain
\begin{equation}\label{Dale_zeta_sigma}
\zeta_n^{(D)}(z)=-\frac{4}{t}\left[\sigma_n^{(D)}(t)+n^2t-\frac{n^2}{2}\right].
\end{equation}

Thus, Dale's determinant solution is expressed in terms of a Painlev\'e VI \(\sigma\)-function with parameters \((0,0,n,n)\). We now show how this relation extends to Cosgrove's generalized \(H_4\) equation with \(h\neq0\).

\subsection{Cosgrove's \(H_4\) equation and Painlev\'e VI}

The generalized \(H_4\) equation \eqref{H_4} can be transformed into the Jimbo-Miwa-Okamoto \(\sigma\)-form of Painlev\'e VI by setting
\begin{equation}\label{H_4-sigma}
H_4(\eta)=\frac{\sigma(t)-\delta^2/2}{t}+\delta^2,\qquad t=\frac{1}{1+\eta}.
\end{equation}
The corresponding Painlev\'e VI parameters satisfy
\begin{equation}
\nu_1^2=\nu_2^2=0,\qquad \nu_3^2=\delta^2+2h,\qquad \nu_4^2=\delta^2-2h.
\end{equation}
For Dale's solution, we have \(\delta=n\) and \(h=0\), giving the parameters \((0,0,n,n)\) obtained in the previous subsection. Taking \(\sigma(t)=\sigma_n^{(D)}(t)\) in \eqref{H_4-sigma} and using \eqref{Dale_zeta_sigma}, we recover
\begin{equation}
\zeta_n^{(D)}(z)=-4H_4(\eta),\qquad z=\eta,
\end{equation}
in agreement with \eqref{Dale-H_4}.

We now consider the generalized case \(h\neq0\) by comparing the Painlev\'e VI parameters associated with Cosgrove's equation and those obtained in \cite{Min:2020ld} for the Hankel determinants \(D_{\Hir}^{(\gamma)}(t;A,B)\). For \(\alpha=\beta=0\), the parameters in \eqref{parameters} reduce to
\begin{equation}
(\nu_1,\nu_2,\nu_3,\nu_4)=(0,0,\Hir,\Hir+\gamma).
\end{equation}
When \(\nu_1=\nu_2=0\), the term \(\nu_1\nu_2\nu_3\nu_4\) vanishes, so the left-hand side of \eqref{sigma_n_equation} no longer depends on the Painlev\'e VI parameters. The right-hand side reduces to \((\sigma')^2(\sigma'+\nu_3^2)(\sigma'+\nu_4^2)\). Thus, to identify the two equations, we only need to compare the factors:
\begin{equation}
(\sigma'+\Hir^2)(\sigma'+(\Hir+\gamma)^2)
\end{equation}
for the Hankel determinant, and
\begin{equation}
(\sigma'+\delta^2+2h)(\sigma'+\delta^2-2h)
\end{equation}
for Cosgrove's equation. For the two differential equations to coincide, these quadratic polynomials in \(\sigma'\) must be identical. Expanding both products and comparing the coefficients of \(\sigma'\) and the constant terms, we obtain
\begin{equation}\label{H_4_parameters}
\Hir^2+(\Hir+\gamma)^2=2\delta^2,\qquad \Hir^2(\Hir+\gamma)^2=\delta^4-4h^2.
\end{equation}
The first relation determines \(\delta^2\) in terms of \(\Hir\) and \(\gamma\). Substituting it into the second relation gives
\begin{equation}\label{parameter_relations}
\delta^2=\frac{\Hir^2+(\Hir+\gamma)^2}{2},\qquad |h|=\frac{\gamma(2\Hir+\gamma)}{4}.
\end{equation}
Here \(\Hir\) is a positive integer, while \(\gamma\geq0\) is the exponent of the weight. The sign of \(h\) is not determined because Cosgrove's \(H_4\) equation depends only on \(h^2\).\footnote{The second equation in Cosgrove's construction, the linear Fuchsian equation for \(K^{(\epsilon)}(\nu,\eta)\) discussed in the Introduction, contains terms proportional to \(h\), unlike the nonlinear equation for \(H_4\), which depends only on \(h^2\). Nevertheless, the equation for \(K^{(\epsilon)}\) is invariant under the simultaneous transformation \((h,\epsilon,\nu)\mapsto(-h,-\epsilon,-\nu)\), where \(\epsilon=\pm1\), as can be verified from Eq.~(3.17) of \cite{Cosgrove:1977np}. Consequently, if \(K_h^{(\epsilon)}(\nu,\eta)\) is a solution for \(h\), then \(K_h^{(\epsilon)}(-\nu,\eta)\) is a solution for \(-h\), with \(\epsilon\) replaced by \(-\epsilon\).} For \(\gamma=0\), we recover \(h=0\) and \(\delta=\Hir\), corresponding to Dale's solution. For \(\gamma>0\), the same construction gives a family of solutions of the generalized \(H_4\) equation with \(h\neq0\). Although \(\gamma\) can take noninteger values, successive Christoffel transformations considered below give integer values of \(\gamma\).

The relations \eqref{parameter_relations} allow us to construct solutions of Cosgrove's generalized \(H_4\) equation using the Hankel determinants studied in \cite{Min:2020ld}. Substituting \eqref{sigma_again} into \eqref{H_4-sigma}, we obtain:
\begin{equation}\label{general_dale_hankel}
H_4(\eta)=(t-1)\frac{d}{dt}\ln\left|D_{\Hir}^{(\gamma)}(t;A,B)\right|+\frac{\gamma^2}{4}\left(2-\frac1t\right),\qquad t=\frac{1}{1+\eta}.
\end{equation}
For \(\gamma=0\), with \(\Hir=n\), \(A=c\), and \(B=1-c\), we recover Dale's original solution. For \(\gamma>0\), \eqref{general_dale_hankel} gives solutions of the generalized \(H_4\) equation with \(h\neq0\), with parameters determined by \eqref{parameter_relations}.

\section{Hankel structure of the Yamazaki-Hori determinants}
\label{section4}

In \cite{Melikyan:2025rmj}, we obtained an equivalent determinant representation of the Yamazaki-Hori solutions of the Ernst equation, involving an arbitrary function \(\kappa\). We recall this representation:
\begin{equation}
\xi_N(x,y;p,q)=\frac{px\,S_N^{(\kappa)}(x,y;p,q\mid x)-iqy\,S_N^{(\kappa)}(x,y;p,q\mid y)}{S_N^{(\kappa)}(x,y;p,q\mid 1)},
\end{equation}
where
\begin{equation}
S_N^{(\kappa)}(x,y;p,q\mid z)=
\begin{vmatrix}
\zeta_0(\kappa) & \zeta_1(\kappa) & \cdots & \zeta_{N-2}(\kappa) & \zeta_{N-1}(\kappa) & b_0 \\
\zeta_1(\kappa) & \zeta_2(\kappa) & \cdots & \zeta_{N-1}(\kappa) & \zeta_N(\kappa) & b_1 \\
\vdots & \vdots & \ddots & \vdots & \vdots & \vdots \\
\zeta_{N-2}(\kappa) & \zeta_{N-1}(\kappa) & \cdots & \zeta_{2N-4}(\kappa) & \zeta_{2N-3}(\kappa) & b_{N-2} \\
\zeta_{N-1}(\kappa) & \zeta_N(\kappa) & \cdots & \zeta_{2N-3}(\kappa) & \zeta_{2N-2}(\kappa) & b_{N-1} \\
a(\kappa)^0 & a(\kappa)^1 & \cdots & a(\kappa)^{N-2} & a(\kappa)^{N-1} & 0
\end{vmatrix}.
\end{equation}
The entries in the last column are:\footnote{
At \(z=0\), this expression is understood by continuity, so that \(b_j^{(\kappa)}(0)=(-\kappa)^j\).
}
\begin{equation}
b_j=b_j^{(\kappa)}(z)=\frac{1}{z}\int_0^z(u^2-\kappa)^j\,du,\qquad j=0,\ldots,N-1,
\end{equation}
while \(a(\kappa)=1-\kappa\). The Hankel matrix elements are
\begin{equation}\label{zeta_arsen}
\zeta_i(\kappa)=\frac{1}{i+1}\left[p^2(x^2-\kappa)^{i+1}+q^2(y^2-\kappa)^{i+1}-(1-\kappa)^{i+1}\right].
\end{equation}
Using \(p^2+q^2=1\), we can express these elements as follows:
\begin{equation}
\zeta_i(\kappa)=p^2\int_{1-\kappa}^{x^2-\kappa}u^i\,du+q^2\int_{1-\kappa}^{y^2-\kappa}u^i\,du.
\end{equation}
For \(p\neq0\), introducing \(s=q/p\), we can write:
\begin{equation}
\zeta_i(\kappa)=p^2\int_{-\infty}^{\infty}u^i w(u;\kappa)\,du,
\end{equation}
where:
\begin{equation}\label{weight_0}
w(u;\kappa)=\theta(x^2-\kappa-u)+s^2\theta(y^2-\kappa-u)-(1+s^2)\theta(1-\kappa-u).
\end{equation}

We now perform the column operations
\begin{equation}
C_j\longmapsto C_j-a(\kappa)C_{j-1},\qquad j=N,N-1,\ldots,2,
\end{equation}
on the first \(N\) columns of \(S_N^{(\kappa)}\). The last row becomes \((1,0,\ldots,0)\), and expansion along this row gives:
\begin{equation}
S_N^{(\kappa)}(x,y;p,q\mid z)=(-1)^{N+2}
\begin{vmatrix}
\widetilde\zeta_0(\kappa) & \widetilde \zeta_1(\kappa) & \cdots & \widetilde\zeta_{N-2}(\kappa) & b_0 \\
\widetilde\zeta_1(\kappa) & \widetilde\zeta_2(\kappa) & \cdots & \widetilde\zeta_{N-1}(\kappa) & b_1 \\
\vdots & \vdots & \ddots & \vdots & \vdots \\
\widetilde\zeta_{N-2}(\kappa) & \widetilde\zeta_{N-1}(\kappa) & \cdots & \widetilde\zeta_{2N-4}(\kappa) & b_{N-2} \\
\widetilde\zeta_{N-1}(\kappa) & \widetilde\zeta_N(\kappa) & \cdots & \widetilde\zeta_{2N-3}(\kappa) & b_{N-1}
\end{vmatrix},
\end{equation}
where
\begin{equation}
\widetilde\zeta_i(\kappa)=\zeta_{i+1}(\kappa)-a(\kappa)\zeta_i(\kappa).
\end{equation}
Using the integral representation of \(\zeta_i(\kappa)\), we obtain
\begin{equation}
\widetilde\zeta_i(\kappa)=p^2\int_{-\infty}^{\infty}u^i\widetilde w(u;\kappa)\,du,
\end{equation}
with the transformed weight
\begin{equation}\label{weight_1}
\widetilde w(u;\kappa)=[u-a(\kappa)]w(u;\kappa).
\end{equation}
This is a Christoffel transformation of the original weight \cite{Ismail-2025oj}. Thus, the column operations reduce the bordered determinant to a smaller determinant whose Hankel block is associated with the weight \(\tilde w\).

We now specialize to \(\kappa=y^2\), as considered in \cite{Melikyan:2025rmj}. In this case, \(a(\kappa)=1-y^2\), and the weight \(w(u;y^2)\) has support in \([0,x^2-y^2]\). We define:
\begin{equation}\label{Definitions}
\Delta:=x^2-y^2,\qquad u=\Delta\xi,\qquad t=\frac{1-y^2}{x^2-y^2}=\frac{1}{1+\eta},\qquad 0<t<1.
\end{equation}
This is the same variable \(t\) introduced earlier for Dale's determinant (cf \eqref{t_def}).

For \(\kappa=y^2\), the weight \eqref{weight_0} becomes:
\begin{equation}
w(u;y^2)=
\begin{cases}
-s^2, & 0<u<1-y^2,\\
1, & 1-y^2<u<\Delta,\\
0, & \text{otherwise}.
\end{cases}
\end{equation}
Since \(a(y^2)=1-y^2=\Delta t\), the transformed weight \eqref{weight_1} can be expressed in terms of the variable \(\xi=u/\Delta\). We obtain:
\begin{equation}
\widetilde\zeta_i(y^2)=p^2\Delta^{i+2}\int_0^1\xi^i\widehat w(\xi,t)\,d\xi,
\end{equation}
where
\begin{equation}\label{weight_3}
\widehat w(\xi,t)=|\xi-t|\left[s^2+(1-s^2)\theta(\xi-t)\right].
\end{equation}
We will also denote this weight by \(w_1(\xi,t)=\widehat w(\xi,t)\).

Consider now the Hankel block of the transformed determinant \(S_N^{(\kappa)}\), evaluated at \(\kappa=y^2\):
\begin{equation}\label{M_hat}
\widehat M_N=
\begin{pmatrix}
\widetilde\zeta_0(y^2) & \widetilde\zeta_1(y^2) & \cdots & \widetilde\zeta_{N-3}(y^2) & \widetilde\zeta_{N-2}(y^2)\\
\widetilde\zeta_1(y^2) & \widetilde\zeta_2(y^2) & \cdots & \widetilde\zeta_{N-2}(y^2) & \widetilde\zeta_{N-1}(y^2)\\
\vdots & \vdots & \ddots & \vdots & \vdots\\
\widetilde\zeta_{N-2}(y^2) & \widetilde\zeta_{N-1}(y^2) & \cdots & \widetilde\zeta_{2N-5}(y^2) & \widetilde\zeta_{2N-4}(y^2)
\end{pmatrix}.
\end{equation}
The weight \eqref{weight_3} has the form \eqref{Chen_weight_again}, with parameters
\begin{equation}
\alpha=\beta=0,\qquad \gamma=1,\qquad A=s^2,\qquad B=1-s^2.
\end{equation}
For real \(p,q\), we have \(A=s^2\geq0\) and \(A+B=1>0\). Thus, unlike Dale's original weight, the transformed weight is nonnegative, and the Painlev\'e VI result of \cite{Min:2020ld} applies after taking the limit \(\alpha,\beta\rightarrow0^+\).

The matrix \(\widehat M_N\) has dimension \((N-1)\times(N-1)\). Comparing its moment representation with \eqref{Hankel_again}, we therefore identify:
\begin{equation}
\Hir=N-1.
\end{equation}
It follows from \eqref{parameters} that the corresponding Painlev\'e VI parameters are
\begin{equation}
(\nu_1,\nu_2,\nu_3,\nu_4)=(0,0,\Hir,\Hir+1)=(0,0,N-1,N).
\end{equation}
Substituting these parameters into \eqref{sigma_n_equation}, we obtain
\begin{equation}
\sigma_{\Hir}'[t(t-1)\sigma_{\Hir}'']^2+\left[2\sigma_{\Hir}'(t\sigma_{\Hir}'-\sigma_{\Hir})-(\sigma_{\Hir}')^2\right]^2
=(\sigma_{\Hir}')^2(\sigma_{\Hir}'+\Hir^2)(\sigma_{\Hir}'+(\Hir+1)^2).
\end{equation}

We now return to the original Hankel matrix elements \eqref{zeta_arsen}, before the column operations. For \(\kappa=y^2\), using the variables introduced in \eqref{Definitions}, we obtain
\begin{equation}\label{zeta_y^2}
\zeta_k(y^2)=p^2\Delta^{k+1}\int_0^1\xi^k w_0(\xi,t;s)\,d\xi,
\end{equation}
where
\begin{equation}\label{sigma_weight}
w_0(\xi,t;s)=-s^2+(1+s^2)\theta(\xi-t).
\end{equation}
This weight coincides with Dale's weight \eqref{Dale's_Weight} for \(c=-s^2\).

Consider the original \(N\times N\) Hankel block
\begin{equation}
M_N(t)=\left(\zeta_{i+j}(y^2)\right)_{i,j=0}^{N-1}.
\end{equation}
Substituting \eqref{zeta_y^2} and extracting the powers of \(\Delta\) and \(p\) from the rows and columns, we obtain
\begin{equation}
\begin{aligned}
\det M_N(t)
&=\Delta^{N^2}p^{2N}\det\left(\int_0^1\xi^{i+j}w_0(\xi,t;s)\,d\xi\right)_{i,j=0}^{N-1}\\
&=\Delta^{N^2}p^{2N}D_N^{(D)}(t,-s^2).
\end{aligned}
\end{equation}
Using the relation between Dale's determinant and its Hankel representation, we therefore find
\begin{equation}\label{Arsen-Dale}
\det M_N(t)=\Delta^{N^2}p^{2N}t^{N^2}A_N(z(t),-s^2).
\end{equation}
Since \(\Delta t=1-y^2\), this can also be written as
\begin{equation}
\det M_N(t)=p^{2N}(1-y^2)^{N^2}A_N(z(t),-s^2).
\end{equation}
Thus, the original Hankel block of the Yamazaki-Hori determinant is directly related to Dale's determinant. In particular, using \eqref{Dale_Hankel}, we obtain
\begin{equation}
\zeta_N^{(D)}(z(t))=4(1-t)\frac{d}{dt}\ln\left|\frac{\det M_N(t)}{p^{2N}\Delta^{N^2}}\right|.
\end{equation}
Since \(t=(1-y^2)/\Delta\) and \(\nu=y/x\) is held fixed, we have \(d\ln\Delta/dt=-\Delta\). Therefore,
\begin{equation}
\zeta_N^{(D)}(z(t))=4(1-t)\left[\frac{d}{dt}\ln|\det M_N(t)|+N^2\Delta\right].
\end{equation}

We have therefore established a direct connection between the original Hankel block of the Yamazaki-Hori determinant and Dale's solution through their common discontinuous weight. For this block, the identification is \(\Hir=N\), giving Painlev\'e VI parameters \((0,0,N,N)\). After the column operations, the remaining Hankel block has order \(\Hir=N-1\), with weight \eqref{weight_3} and parameters \((0,0,N-1,N)\).

\subsection{Relation between the Hankel determinants}

We have identified two Hankel blocks in the determinant representation of the Yamazaki-Hori solutions: the original \(N\times N\) matrix \(M_N\) and the transformed \((N-1)\times(N-1)\) matrix \(\widehat M_N\).
 Their respective weights, given by \eqref{sigma_weight} and \eqref{weight_3}, satisfy:
\begin{equation}
w_1(\xi,t)=(\xi-t)w_0(\xi,t;s).
\end{equation}
Thus, the column operations correspond to a Christoffel transformation of the original weight, while the size of the Hankel block decreases by one. For \(M_N\), we have \(\Hir=N\) and \(\gamma=0\), giving Painlev\'e VI parameters \((0,0,N,N)\). For \(\widehat M_N\), we have \(\Hir=N-1\) and \(\gamma=1\), giving parameters \((0,0,N-1,N)\).

We now examine the relation between the corresponding Painlev\'e VI \(\sigma\)-functions. Using the moment representation of \(\widehat M_N\), we obtain
\begin{equation}\label{D_hat}
\widehat D_N(t):=\det\widehat M_N=p^{2(N-1)}\Delta^{N(N-1)}D_{N-1}^{(1)}(t;s^2,1-s^2).
\end{equation}
To compare the determinants at the same Hankel order \(n\), we take \(M_n\) from the Yamazaki-Hori solution of order \(n\) and \(\widehat M_{n+1}\) from the solution of order \(n+1\). Both matrices have dimension \(n\times n\). Using \eqref{Arsen-Dale} and \eqref{D_hat}, we obtain
\begin{equation}
\begin{aligned}
\det M_n &=p^{2n}\Delta^{n^2}D_n^{(D)}(t,-s^2),\\
\det\widehat M_{n+1} &=p^{2n}\Delta^{n(n+1)}D_n^{(1)}(t;s^2,1-s^2).
\end{aligned}
\end{equation}
We define their ratio by
\begin{equation}\label{Y_n}
Y_n(t):=\frac{\det\widehat M_{n+1}}{\det M_n}=\Delta^n\frac{D_n^{(1)}(t;s^2,1-s^2)}{D_n^{(D)}(t,-s^2)}.
\end{equation}
To compare the corresponding Painlev\'e VI \(\sigma\)-functions, we define:
\begin{equation}\label{eq:R_n}
R_n(t):=\frac{Y_n(t)}{\Delta^n}=\frac{D_n^{(1)}(t;s^2,1-s^2)}{D_n^{(D)}(t,-s^2)}.
\end{equation}

For \(\gamma=0\), the corresponding Painlev\'e VI \(\sigma\)-function is given by \eqref{sigma_Dale}. For \(\gamma=1\), using \eqref{sigma_again} with \(\alpha=\beta=0\), we obtain
\begin{equation}
\begin{aligned}
\sigma_n^{(0)}(t)&=t(t-1)\frac{d}{dt}\ln|D_n^{(D)}(t,-s^2)|-n^2t+\frac{n^2}{2},\\
\sigma_n^{(1)}(t)&=t(t-1)\frac{d}{dt}\ln D_n^{(1)}(t;s^2,1-s^2)-n(n+1)t+\frac{n(n+1)}{2}.
\end{aligned}
\end{equation}
Taking the logarithmic derivative of \eqref{eq:R_n}, we obtain
\begin{equation}
\frac{d}{dt}\ln|R_n(t)|=\frac{d}{dt}\ln D_n^{(1)}(t;s^2,1-s^2)-\frac{d}{dt}\ln|D_n^{(D)}(t,-s^2)|.
\end{equation}
Subtracting the two expressions for the \(\sigma\)-functions, we therefore obtain
\begin{equation}\label{sigma_relation}
\sigma_n^{(0)}(t)=\sigma_n^{(1)}(t)-t(t-1)\frac{d}{dt}\ln|R_n(t)|+n\left(t-\frac12\right).
\end{equation}
This relation expresses the difference between the two Painlev\'e VI \(\sigma\)-functions, with parameters \((0,0,n,n)\) and \((0,0,n,n+1)\), in terms of the logarithmic derivative of the ratio of the original and Christoffel-transformed Hankel determinants.

\section{Nested Hankel structures and Painlev\'e VI parameters}\label{section5}

In the previous sections, we showed that the first Christoffel transformation of the Yamazaki-Hori Hankel block changes the corresponding Painlev\'e VI parameters from \((0,0,N,N)\) to \((0,0,N-1,N)\). We now consider successive transformations and determine the Hankel structures and Painlev\'e VI parameters obtained at each step.

\begin{theorem}
Let \(N\) denote the order of the original Yamazaki-Hori solution, and consider its determinant representation with \(\kappa=y^2\) and \(p\neq0\). After \(k\) successive Christoffel transformations and the corresponding reductions of the Hankel blocks, with \(0\leq k\leq N-1\), the remaining block has dimension \(N-k\), and its weight has exponent \(\gamma=k\). The associated Painlev\'e VI parameters are
\begin{equation}\label{nested_parameters}
(\nu_1,\nu_2,\nu_3,\nu_4)=(0,0,N-k,N).
\end{equation}
\end{theorem}
\begin{proof}
We use the choice \(\kappa=y^2\) introduced in the preceding section, for which \(a=1-y^2=\Delta t\). The original moments are
\begin{equation}
\zeta_j^{(0)}:=\zeta_j(y^2)=p^2\int_0^\Delta u^j w(u;y^2)\,du.
\end{equation}
The first Christoffel transformation gives
\begin{equation}\label{eq:first-transformed-moment}
\zeta_j^{(1)}=\zeta_{j+1}^{(0)}-a\zeta_j^{(0)}.
\end{equation}
Substituting the integral representation of the original moments, we obtain
\begin{equation}
\zeta_j^{(1)}=p^2\int_0^\Delta u^j(u-a)w(u;y^2)\,du.
\end{equation}
Thus, the first transformation corresponds to multiplication of the original weight by \(u-a\), as established in Section~\ref{section4}.

We can apply the same transformation to the new moments. At the second step,
\begin{equation}\label{eq:second-recursion}
\zeta_j^{(2)}=\zeta_{j+1}^{(1)}-a\zeta_j^{(1)}.
\end{equation}
Using \eqref{eq:first-transformed-moment}, we obtain
\begin{equation}\label{eq:second-expanded}
\zeta_j^{(2)}=\zeta_{j+2}^{(0)}-2a\zeta_{j+1}^{(0)}+a^2\zeta_j^{(0)},
\end{equation}
and therefore
\begin{equation}\label{eq:second-integral}
\zeta_j^{(2)}=p^2\int_0^\Delta u^j(u-a)^2w(u;y^2)\,du.
\end{equation}
Thus, the two successive transformations multiply the original weight by \((u-a)^2\), corresponding to a second-order Christoffel transformation.

Repeating this procedure, we obtain:
\begin{equation}\label{eq:general-recursion}
\zeta_j^{(r+1)}=\zeta_{j+1}^{(r)}-a\zeta_j^{(r)},\qquad r\geq0.
\end{equation}
The above calculations show that each transformation introduces one additional factor \(u-a\). By induction,
\begin{equation}\label{eq:general-integral}
\zeta_j^{(r)}=p^2\int_0^\Delta u^j(u-a)^r w(u;y^2)\,du.
\end{equation}

We next examine the Hankel structure at an arbitrary step \(r\). Let \(m=N-r\) be the dimension of the remaining block, which we denote by:
\begin{equation}\label{eq:generic-nested-block}
M_m^{(r)}=
\begin{pmatrix}
\zeta_0^{(r)} & \zeta_1^{(r)} & \cdots & \zeta_{m-1}^{(r)}\\
\zeta_1^{(r)} & \zeta_2^{(r)} & \cdots & \zeta_m^{(r)}\\
\vdots & \vdots & \ddots & \vdots\\
\zeta_{m-1}^{(r)} & \zeta_m^{(r)} & \cdots & \zeta_{2m-2}^{(r)}
\end{pmatrix}.
\end{equation}
The entries depend only on the sum of their row and column indices, so the Hankel structure is preserved at every step.

Applying the column operations:
\begin{equation}\label{eq:iterated-column-operations}
C_j\longmapsto C_j-aC_{j-1},\qquad j=m,m-1,\ldots,2,
\end{equation}
and using \eqref{eq:general-recursion}, we obtain
\begin{equation}\label{eq:nested-matrix-step}
\left(
\begin{array}{c|cccc}
\zeta_0^{(r)} & \zeta_0^{(r+1)} & \zeta_1^{(r+1)} & \cdots & \zeta_{m-2}^{(r+1)} \\
\zeta_1^{(r)} & \zeta_1^{(r+1)} & \zeta_2^{(r+1)} & \cdots & \zeta_{m-1}^{(r+1)} \\
\vdots & \vdots & \vdots & \ddots & \vdots \\
\zeta_{m-2}^{(r)} & \zeta_{m-2}^{(r+1)} & \zeta_{m-1}^{(r+1)} & \cdots & \zeta_{2m-4}^{(r+1)} \\ \noalign{\vskip 2pt}
\hline
\noalign{\vskip 2pt}
\zeta_{m-1}^{(r)} & \zeta_{m-1}^{(r+1)} & \zeta_m^{(r+1)} & \cdots & \zeta_{2m-3}^{(r+1)}
\end{array}
\right).
\end{equation}

The vertical and horizontal lines separate the first column and the last row from the remaining entries. Removing this column and row leaves the \((m-1)\times(m-1)\) Hankel block
\begin{equation}\label{eq:next-nested-block}
M_{m-1}^{(r+1)}=
\begin{pmatrix}
\zeta_0^{(r+1)} & \zeta_1^{(r+1)} & \cdots & \zeta_{m-2}^{(r+1)}\\
\zeta_1^{(r+1)} & \zeta_2^{(r+1)} & \cdots & \zeta_{m-1}^{(r+1)}\\
\vdots & \vdots & \ddots & \vdots\\
\zeta_{m-2}^{(r+1)} & \zeta_{m-1}^{(r+1)} & \cdots & \zeta_{2m-4}^{(r+1)}
\end{pmatrix}.
\end{equation}
Thus, the successive nested blocks form the sequence
\begin{equation}\label{eq:nested-hierarchy}
\begin{aligned}
M_N^{(0)}
&\longrightarrow M_{N-1}^{(1)}
\longrightarrow M_{N-2}^{(2)}
\longrightarrow\cdots \longrightarrow M_{N-r}^{(r)}
\longrightarrow\cdots
\longrightarrow M_1^{(N-1)}.
\end{aligned}
\end{equation}
In particular, \(M_N^{(0)}=M_N\) and \(M_{N-1}^{(1)}=\widehat M_N\). Each step consists of a Christoffel transformation of the moments followed by the selection of a smaller Hankel block. The corresponding weights are obtained as follows. Using \(u=\Delta\xi\) and \(a=\Delta t\) in \eqref{eq:general-integral}, we have:
\begin{equation}
\zeta_j^{(r)}=p^2\Delta^{j+r+1}\int_0^1\xi^j w_r(\xi,t;s)\,d\xi,
\end{equation}
where
\begin{equation}\label{eq:general-weight}
w_r(\xi,t;s)=(\xi-t)^r w_0(\xi,t;s).
\end{equation}
Since the original weight is given by \eqref{sigma_weight}, we find
\begin{equation}\label{eq:general-piecewise}
w_r(\xi,t;s)=
\begin{cases}
(-1)^{r+1}s^2|\xi-t|^r, & 0<\xi<t,\\
|\xi-t|^r, & t<\xi<1.
\end{cases}
\end{equation}
Equivalently,
\begin{equation}
w_r(\xi,t;s)=|\xi-t|^r\left[A_r+B_r\theta(\xi-t)\right],
\end{equation}
where
\begin{equation}
A_r=(-1)^{r+1}s^2,\qquad B_r=1-(-1)^{r+1}s^2.
\end{equation}
Comparing this expression with \eqref{Chen_weight_again}, we identify
\begin{equation}
\alpha=\beta=0,\qquad \gamma=r.
\end{equation}

We first analyze the signs of the weights \eqref{eq:general-piecewise}. For real \(p,q\) with \(pq\neq0\), the original weight is negative for \(0<\xi<t\) and positive for \(t<\xi<1\). Since each Christoffel transformation multiplies the weight by \(\xi-t\), the weights \(w_r\) are nonnegative when \(r\) is odd and change sign when \(r\) is even. For \(q=0\), all weights are nonnegative. At odd levels, we have \(A_r=s^2>0\) and \(A_r+B_r=1>0\). Thus, the positivity conditions of \cite{Min:2020ld} are satisfied, and the corresponding Painlev\'e VI result applies after taking the limit \(\alpha,\beta\rightarrow0^+\).

For even \(r\geq2\), the weights change sign, so the positivity conditions of \cite{Min:2020ld} are not satisfied. We therefore use the analytic continuation \eqref{continuation_Manko}, under which \(p=i\tilde p\) and \(q^2=1+\tilde p^2\). Consequently, \(s^2=-1-1/\tilde p^2\), so that \(A_r=-s^2=1+1/\tilde p^2>0\), while \(A_r+B_r=1>0\). The continued weight is therefore nonnegative throughout \((0,1)\), and the corresponding Painlev\'e VI result applies in this region, again taking the limit \(\alpha,\beta\rightarrow0^+\).

To extend this result to the original real values of \(p,q\), we consider the moments of \(w_r\), which for each fixed \(r\) are given by:
\begin{equation}
\int_0^1\xi^j w_r(\xi,t;s)\,d\xi=(-1)^{r+1}s^2\int_0^t\xi^j(t-\xi)^r\,d\xi+\int_t^1\xi^j(\xi-t)^r\,d\xi.
\end{equation}
Thus, the moments depend linearly on \(s^2\), and the corresponding Hankel determinant and its derivatives with respect to \(t\) are polynomials in \(s^2\). After substituting the logarithmic derivatives into the Painlev\'e VI equation and clearing denominators, we obtain a polynomial identity in \(s^2\). Since this identity holds for the continued values \(s^2<-1\), it holds identically and therefore remains valid for the original values \(s^2\geq0\), wherever the Hankel determinant is nonzero. The case \(r=0\) was already established in Section~\ref{section3}, using \cite{Chen:2010rl}.

The block \(M_{N-r}^{(r)}\) has dimension \(N-r\). Substituting the moment representation into its determinant and extracting the factors of \(p^2\) and \(\Delta\) from the rows and columns, we obtain:
\begin{equation}\label{eq:nested-determinant}
\det M_{N-r}^{(r)}
=p^{2(N-r)}\Delta^{N(N-r)}
D_{N-r}^{(r)}(t;A_r,B_r).
\end{equation}
Thus, the Hankel determinant on the right has order \(\Hir=N-r\) and weight exponent \(\gamma=r\). Using \eqref{parameters}, the corresponding Painlev\'e VI parameters are
\begin{equation}
\begin{aligned}
(\nu_1,\nu_2,\nu_3,\nu_4)
&=(0,0,\Hir,\Hir+\gamma)\\
&=(0,0,N-r,N).
\end{aligned}
\end{equation}
Setting \(r=k\) proves \eqref{nested_parameters}.
\end{proof}

The hierarchy also allows us to relate the Painlev\'e VI \(\sigma\)-functions corresponding to successive Hankel blocks. At iteration \(r\), we have \(\Hir=N-r\) and \(\gamma=r\). Using \eqref{c1_again} and \eqref{c2_again}, the constants entering \eqref{sigma_again} become
\begin{equation}
c_1^{(r)}=-N(N-r),\qquad c_2^{(r)}=\frac{N(N-r)}{2}.
\end{equation}
Consequently, the corresponding \(\sigma\)-function is
\begin{equation}\label{eq:nested-sigma}
\sigma_{N-r}^{(r)}(t)=t(t-1)\frac{d}{dt}\ln\left|D_{N-r}^{(r)}(t;A_r,B_r)\right|-N(N-r)\left(t-\frac12\right).
\end{equation}
The next block has order \(N-r-1\) and weight exponent \(r+1\). To compare the two \(\sigma\)-functions, define
\begin{equation}\label{eq:nested-ratio}
\mathcal{Y}_{N,r}(t):=\frac{D_{N-r-1}^{(r+1)}(t;A_{r+1},B_{r+1})}{D_{N-r}^{(r)}(t;A_r,B_r)}=p^2\Delta^N\frac{\det M_{N-r-1}^{(r+1)}}{\det M_{N-r}^{(r)}}.
\end{equation}
Subtracting \eqref{eq:nested-sigma} from the corresponding expression at iteration \(r+1\), we obtain
\begin{equation}\label{eq:nested-sigma-relation}
\sigma_{N-r}^{(r)}(t)=\sigma_{N-r-1}^{(r+1)}(t)-t(t-1)\frac{d}{dt}\ln|\mathcal{Y}_{N,r}(t)|-N\left(t-\frac12\right).
\end{equation}
This relation connects consecutive Painlev\'e VI \(\sigma\)-functions with parameters \((0,0,N-r,N)\) and \((0,0,N-r-1,N)\).

Applying \eqref{eq:nested-sigma-relation} successively, we can relate the original \(\sigma\)-function to any level of the hierarchy:
\begin{equation}\label{eq:nested-sigma-general}
\begin{aligned}
\sigma_N^{(0)}(t)={}&\sigma_{N-k}^{(k)}(t)-t(t-1)\frac{d}{dt}\ln\left|\frac{D_{N-k}^{(k)}(t;A_k,B_k)}{D_N^{(0)}(t;A_0,B_0)}\right|-Nk\left(t-\frac12\right),
\end{aligned}
\end{equation}
where \(0\leq k\leq N-1\). In particular, for \(k=N-1\), this relation connects the original \(\sigma\)-function to that of the final Hankel block of order one. More generally, \eqref{eq:nested-sigma-general} relates the \(\sigma\)-functions at different levels of the hierarchy through ratios of their normalized Hankel determinants.

At each step, the third Painlev\'e VI parameter decreases by one, while the fourth remains equal to \(N\). Using \eqref{parameter_relations}, the corresponding parameters of Cosgrove's generalized \(H_4\) equation are:

\begin{equation}\label{eq:nested-cosgrove-parameters}
\delta^2=\frac{(N-k)^2+N^2}{2},\qquad |h|=\frac{k(2N-k)}{4}.
\end{equation}
For \(k=0\), we recover the original case \(h=0\), with \(\delta=N\). For \(k\geq1\), the associated Hankel determinants give solutions of Cosgrove's generalized equation with \(h\neq0\).

The hierarchy terminates after \(N-1\) transformations, when the Hankel block reduces to a single matrix element. It is natural to ask whether the original Hankel structure can be recovered by reversing these transformations. Although the weights are explicitly related by \(w_r(\xi,t;s)=(\xi-t)^r w_0(\xi,t;s)\), the final block contains only one moment, which is not sufficient to reconstruct the original matrix. Additional information is therefore needed to reverse the hierarchy.

A related question is whether this reconstruction could lead directly to the Yamazaki-Hori solutions of the Ernst equation \cite{Yamazaki:1977xd, Hori:1978ie}. Cosgrove's construction involves two differential equations: a nonlinear equation for \(H_4(\eta)\), related to Painlev\'e VI, and a linear equation for \(K^{(\epsilon)}(\nu,\eta)\). We have established a connection between the first equation and the Hankel blocks appearing in the numerator and denominator of the Yamazaki-Hori solutions. It remains to understand whether the second equation can account for the border entries of these determinants. Establishing this connection, together with a reconstruction of the Hankel hierarchy, could provide a way to derive the Yamazaki-Hori solutions directly from Cosgrove's two equations, rather than starting from their known determinant representation.

\section{Conclusion}

We have established a relation between Dale's determinant solutions of Cosgrove's equation and the Hankel structures appearing in the Yamazaki-Hori solutions of the Ernst equation. Using Cosgrove's asymptotic condition, we determined \(c=-q^2/p^2\) and showed that Dale's determinant can be expressed in terms of a discontinuous weight. This allowed us to identify the corresponding Painlev\'e VI parameters and obtain Hankel determinant solutions of Cosgrove's generalized \(H_4\) equation, including the case \(h\neq0\).

We also showed that the original Yamazaki-Hori Hankel block is directly related to Dale's determinant. Successive Christoffel transformations generate a nested hierarchy of Hankel blocks, with each transformation reducing the order by one. Starting from order \(N\), the block obtained after \(k\) transformations corresponds to Painlev\'e VI parameters \((0,0,N-k,N)\). We also derived relations between the associated Painlev\'e VI \(\sigma\)-functions.

An interesting question is whether the original Hankel structure can be reconstructed by reversing these transformations. Another is whether Cosgrove's second, linear differential equation can account for the border entries of the Yamazaki-Hori determinants. Establishing these connections could provide a way to derive the Yamazaki-Hori solutions directly from Cosgrove's two equations.

\bibliographystyle{elsarticle-num}
\bibliography{pvifinal}
\end{document}